\documentclass{article}
\usepackage{spconf,amsmath,graphicx}
\usepackage{multirow}

\title{Using Signal Processing in Tandem With Adapted Mixture Models for Classifying Genomic Signals}
\name{Saish Jaiswal$^1$, Shreya Nema$^2$, Hema A Murthy$^1$, Manikandan Narayanan$^{1,3,4}$}
\address{$^1$Department of Computer Science and Engineering, Indian Institute of Technology (IIT) Madras\\
$^2$Department of Biotechnology, IIT Madras\\$^3$Center for Integrative Biology and Systems Medicine, IIT Madras\\$^4$Robert Bosch Centre for Data Science and Artificial Intelligence, IIT Madras}

\begin{document}

\maketitle

\begin{abstract}
Genomic signal processing has been used successfully in bioinformatics to analyze biomolecular sequences and gain varied insights into DNA structure, gene organization, protein binding, sequence evolution, etc. But challenges remain in finding the appropriate spectral representation of a biomolecular sequence, especially when multiple variable-length sequences need to be handled consistently. In this study, we address this challenge in the context of the well-studied problem of classifying genomic sequences into different taxonomic units (strain, phyla, order, etc.). We propose a novel technique that employs signal processing in tandem with Gaussian mixture models to improve the spectral representation of a sequence and subsequently the taxonomic classification accuracies. The sequences are first transformed into spectra, and projected to a subspace, where sequences belonging to different taxons are better distinguishable. Our method outperforms a similar state-of-the-art method on established benchmark datasets by an absolute margin of 6.06\% accuracy.
\end{abstract}

\section{Introduction}
\label{sec:intro}
    Any sequence of symbols can be mapped to a time series or signal. Biomolecular or genomic sequences, such as DNA or protein sequences that encode information required for the functioning of a cell, can also be viewed as signals and processed for gaining insights about living organisms. This genomic signal processing (GSP) perspective has been appreciated since the early 2000s \cite{anastassiou2001genomic, vaidyanathan2004genomics}, where space-varying signal across the length of the sequence has been utilized to understand properties of DNA, RNA, and protein sequences. For instance, GSP has been employed for varied tasks in molecular biology, including identifying protein-coding regions in DNA sequences, studying functional domains or binding sites in protein sequences, or more recently the classification of genome sequences into different taxonomic categories (such as species, genus, family, order, etc.).   
    
    Obtaining a systematic frequency or spectral representation of each biomolecular sequence is a key step in any of these GSP applications. Within the context of taxonomic classification of sequences, authors of a GSP study \cite{skutkova2013classification} for instance used dynamic time warping on variable-length sequence representations to classify genomic signals. But, the study is limited to DNA data from ten organisms and the length of the sequences did not vary much. Authors of two more recent studies \cite{randhawa2019ml, randhawa2020machine} utilized GSP along with machine learning to classify genomic sequences at different taxonomic levels, and more importantly, could handle sequences of widely-varying lengths. But they do so by length-normalization (i.e., truncating a long sequence or padding a short sequence to a fixed length), which can possibly lead to loss or distortion of information in the fixed-length spectral representation of variable-length sequences.
    
    In this work\footnote{This work was supported by the Wellcome Trust/DBT India Alliance Intermediate Fellowship Grant IA/I/17/2/503323 awarded to M.N.}, we propose a novel feature extraction technique that employs GSP and Gaussian mixture models to get a fixed-length subspace representation for genome sequences of varying lengths corresponding to different species. Unlike the existing methods \cite{randhawa2019ml, randhawa2020machine} which rely on truncation/padding of DNA sequences, we take the entire sequence information into consideration to compute spectral features. More specifically, we use a sliding window of fixed length to highlight the spectral variations across the length of the sequence, and use a ``Universal Background Model (UBM)'' based Gaussian Mixture Model (GMM) to project these representations into a fixed-dimensional subspace. Our technique enables different mixtures to flexibly capture the varying properties of the sequences, and thereby obtain informative subspace representations, which can then be used for better taxonomic classification. We consider the earlier GSP study \cite{randhawa2019ml} as the state-of-the-art baseline to assess the benefits of our approach.

    \begin{center}
        \begin{figure*}[!ht]
            \centering
            \includegraphics[width=\linewidth]{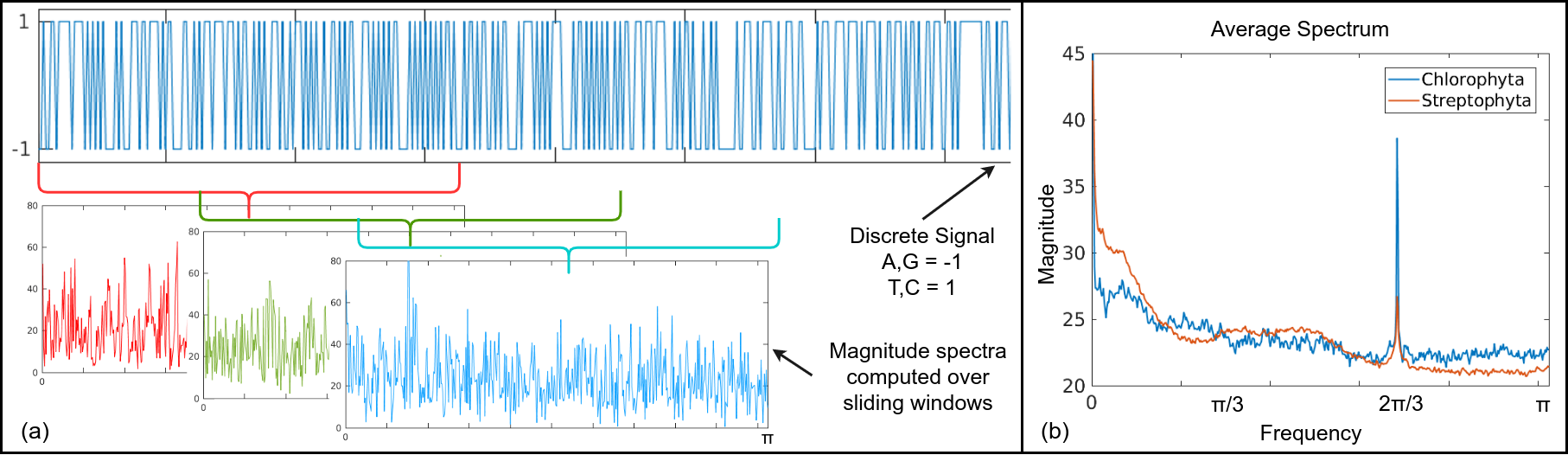}
            \caption{(a) Extraction of magnitude spectra from sliding windows. (b) Average of spectra computed over sequences of the Plant class shows discrimination between the two sub-classes (see Figure S1 to S6 in Suppl. Material for other classes). We employ UBM-GMM to improve upon these average representations.}
            \label{fig:SpectralAnalysis}
        \end{figure*}
    \end{center}
    
\section{Background}
\label{sec:background}

    \subsection{Genomic Signal Representation}
    Genomic (DNA) sequences can be converted to discrete signals using any numerical representation, where each of the bases is associated with a numerical value. Digital signal processing can then be applied to these discrete signals. We can assign some numerical values to each of the nucleotides in a DNA sequence without using any biological knowledge, or by using some physical or chemical properties of the nucleotides \cite{kwan2009numerical, borrayo2014genomic, adetiba2016classification, adetiba2016identification}. In the baseline study \cite{randhawa2019ml}, 13 different numerical representations were tried, and we also adopt a similar strategy; but we present results for one of the best-performing representations viz., the Purine-Pyrimidine representation (purines (A, G) = -1 and pyrimidines (T, C) = 1)).

    \subsection{Feature Extraction and Taxonomic Classification in the Baseline GSP Method}
    \label{sec:featureExtraction}
    Digital signal processing enables the study of the spectral properties of sequences. Distinct spectral properties enable classification. In the baseline, study \cite{randhawa2019ml}, the sequences of different lengths are normalized (i.e., truncated or  ``anti-symmetrically'' padded) to the median length of all sequences. Discrete Fourier Transform (DFT) is then applied to the length-normalized sequences to extract the corresponding magnitude spectra. The fixed-length spectra were then used to compute pairwise distance matrices using both Pearson's Correlation Coefficient (PCC) and Euclidean distance metrics. The corresponding columns of the distance matrix were considered as feature vectors for each of the sequences for the classification task.

    \begin{center}
        \begin{figure*}[!ht]
            \centering
            \includegraphics[width=\linewidth]{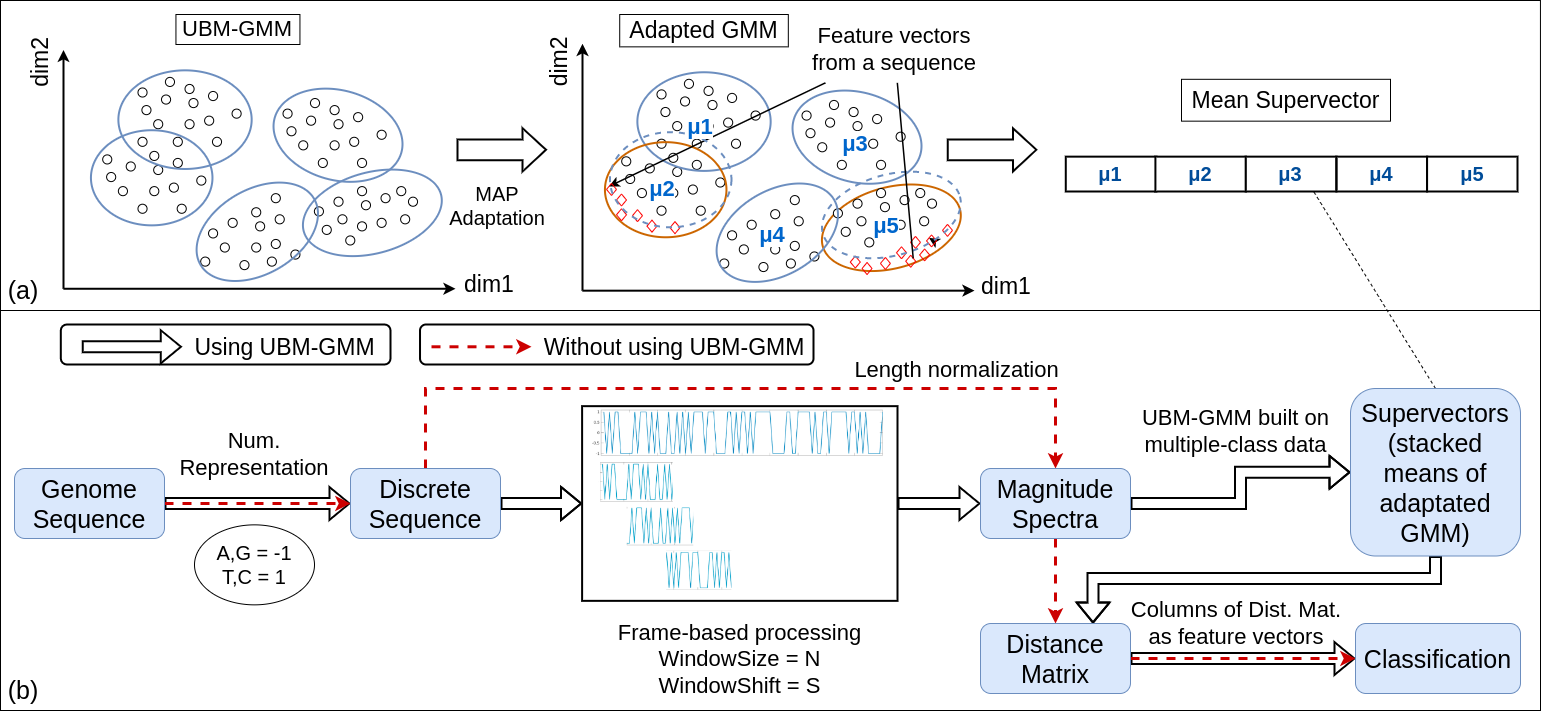}
            \caption{(a) Adapting GMM from UBM-GMM to get a sub-space representation. (b) The figure depicts the baseline approach (without UBM-GMM) as well as the UBM-GMM-based approach. In the latter approach, we employ spectra computed over sliding windows to get a fixed dimensional mean-supervector representation.}
            \label{fig:Pipeline}
        \end{figure*}
    \end{center}

    \subsection{Universal Background Model}
    \label{sec:UBM-GMM}
    A Universal Background Model (UBM) is a Gaussian mixture model (GMM) that has been widely used for speaker recognition tasks \cite{reynolds2000speaker}. We use this model in our study to obtain a fixed-length subspace representation for genomic sequences. A GMM is a generative model trained on data belonging to one particular class. A huge amount of data is required to estimate the parameters of a GMM accurately, especially when the dimension of the feature vector is quite large. UBM-GMM is employed in such cases to tackle the issue of data availability. It is built on data pooled from different classes and hence it tries to capture the properties of all of these classes. Moreover, owing to the availability of a large amount of data, the parameters of UBM-GMM are robustly estimated. Then a particular class's data can be used to adapt a GMM from the trained UBM-GMM using maximum-a-posteriori adaptation (MAP).

    \section{Proposed Work}
    \label{sec:experimentalDesign}
    
    \subsection{Our UBM-GMM-based subspace representation}
    We propose a novel UBM-GMM-based approach to obtain a fixed-length subspace representation for variable-length sequences, which can then be used for the genome classification task. The main contribution of our approach is that we can handle sequences of widely-varying lengths, without suffering the likely loss of information associated with length normalization (padding/truncation) of sequences done in existing studies  \cite{randhawa2019ml,randhawa2020machine}. This loss of information is a serious issue when the lengths of the sequences vary quite a lot (which is the case for many classes in our application (Tables \ref{tab:Dataset},\ref{tab:Accuracy})). 
    
    The key idea behind our approach is that we take into account information across the entire sequence, by extracting features from every frame (sliding window) of the sequence (Figure \ref{fig:SpectralAnalysis} (a)). This yields multiple spectra from a given sequence, and we pool all such spectra corresponding to all the sequences across all the classes to build a UBM-GMM (see Figure \ref{fig:Pipeline}) and subsequently get a fixed-length representation. 
    
    To provide more detail, we slide a window (with overlap across windows) of fixed length across the sequence, where we extract spectral features for every segment of the sequence. As discussed in \cite{vaidyanathan2004genomics}, we experimented with a sufficiently large window of sizes ranging from 351 to a few thousand nucleotides. We considered window shifts ranging from 100 to 300 nucleotides. The magnitude spectrum is computed for each windowed sequence, with the DFT order chosen appropriately (window size rounded up to the nearest power of two). These magnitude spectra were then used to build a UBM which projects the spectra to a fixed-dimensional subspace, while primarily ensuring that the variability of the sequences across the entire sequence length is captured in the representation. The number of mixture components for UBM-GMM was empirically chosen as ten.
    
    When adapting a GMM from the UBM-GMM using sequence data, we do mean-only adaptation using MAP, i.e., only the means of the UBM-GMM are changed depending on the feature vectors of the given sequence. The covariance matrices of the mixture components are not adapted due to insufficient data points from a sequence. The means of the adapted GMM are stacked to get a mean-supervector (Figure \ref{fig:Pipeline}), which acts as the fixed-length subspace representation vector for the entire (untruncated/unpadded) sequence. 

    \subsection{Taxonomic classification using the UBM-GMM-based representation}
    Once the subspace representations of sequences are obtained using our novel UBM-GMM-based approach, we can use the rest of the setup from the baseline approach for the taxonomic classification task, in order to perform a fair comparison of our and baseline representations' classification accuracy. Specifically, our subspace representations are used to calculate a pair-wise distance matrix using PCC (as this was found to be the better metric in \cite{randhawa2019ml}), and the distance matrix columns used as feature vectors for the classification task. We used four different machine learning classifiers — Linear Discriminant, Linear Support Vector Machine (SVM), Quadratic SVM, and K-Nearest Nighbours (KNN) — as in the baseline study  \cite{randhawa2019ml}.


    \begin{center}
        \begin{figure*}[!h]
            \centering
            \includegraphics[width=\linewidth]{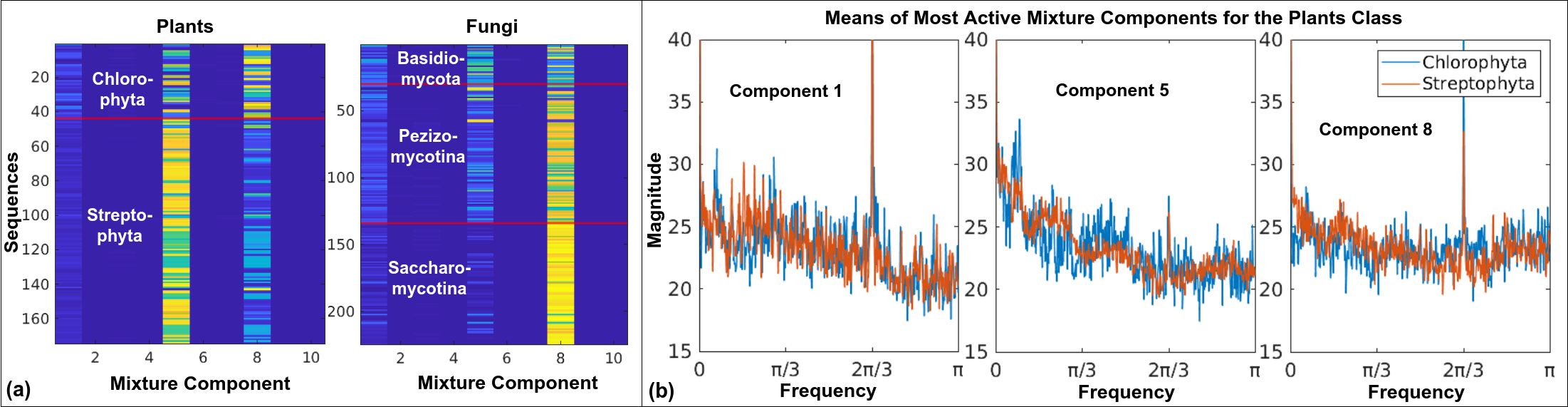}
            \caption{(a) Mixture activations while adapting 10-mixture-component UBM-GMM, (b) Means of most active components.}
            \label{fig:PPMC}
        \end{figure*}
    \end{center}

\vspace{-4em}
\section{Experimentation}
\label{sec:experiment}

    \subsection{Dataset}
    \label{sec:dataset}
    All the datasets used in this study are available from the National Center for Biotechnology Information (NCBI) and through links provided in the Suppl. material \footnote{Supplementary material/codes: https://sites.google.com/view/genomic-signal-processing/}. The datasets consist of mitochondrial and whole genome sequences corresponding to 8 classes. Each of these classes consists of two or more sub-classes, and classification is done at the sub-class level.
    
    \begin{table}[!h]
    \centering
    \caption{Benchmark dataset details shown for certain classes here (see Table S1 in Suppl. Material for remaining classes)}
    \vspace{2mm}
    \label{tab:Dataset}
    \scalebox{0.85}{
    \begin{tabular}{|l|l|l|} 
    \hline
    \multicolumn{1}{|c|}{Class}                            & \multicolumn{1}{c|}{Subclasses (no. of seq.)}                                                                                                                         & \multicolumn{1}{c|}{Statistics}                                                \\ 
    \hline
    Fungi                                                  & \begin{tabular}[c]{@{}l@{}}Basidiomycota (30),\\Pezizomycotina (104),\\Saccharomycotina (92)\end{tabular}                                                             & \begin{tabular}[c]{@{}l@{}}Min: 1364; Max: 235849\\Med: 39154\end{tabular}     \\ 
    \hline
    Plants                                                 & \begin{tabular}[c]{@{}l@{}}Chlorophyta (44),\\Streptophyta (130)\end{tabular}                                                                                         & \begin{tabular}[c]{@{}l@{}}Min: 12998; Max: 1999595\\Med: 128211\end{tabular}  \\ 
    \hline
    \end{tabular}
    }
    \end{table}

\subsection{Results and Analysis}
We ran the experiments on the benchmark datasets using the baseline as well as our approach. We experimented with different numerical representations, distance metrics to compute pair-wise distance matrices, window sizes, and window shifts. Based on the performances, we used the Purine-Pyrimidine representation, median length normalization (for the baseline approach), Pearson's correlation coefficient (PCC) as a distance metric, window size as 702, window shift as 300 and FFT Order as 1024. The average accuracies computed over earlier discussed four different classifiers are mentioned in Table \ref{tab:Accuracy}.

\begin{table}[!h]
\centering
\caption{Comparison of the average accuracies over four different classifiers. For details on individual accuracies of each classifier, see Table S2 in the Suppl. material.}
\vspace{2mm}
\label{tab:Accuracy}
\scalebox{0.85}{
\begin{tabular}{|l|c|c|c|c|} 
\hline
\multicolumn{1}{|c|}{\multirow{2}{*}{Class}} & \multicolumn{2}{c|}{Average Accuracy} & \multirow{2}{*}{\begin{tabular}[c]{@{}c@{}}Median seq.\\length\end{tabular}} & \multirow{2}{*}{MAD}  \\ 
\cline{2-3}
\multicolumn{1}{|c|}{}                       & Baseline        & Our approach        &                                                                              &                       \\ 
\hline
Primates                                     & 98.975          & 100                 & 16554                                                                        & 50                    \\ 
\hline
Protists                                     & 78.775          & 95.75               & 35660                                                                        & 13944                 \\ 
\hline
Fungi                                        & 79.925          & 91.4                & 39154                                                                        & 34768                 \\ 
\hline
Plants                                       & 89.65           & 97.275              & 128211                                                                       & 290801                \\ 
\hline
Insects                                      & 89.95           & 96.975              & 15529                                                                        & 768                   \\ 
\hline
Vertebrates                                  & 98.55           & 99.7                & 16616                                                                        & 697                   \\ 
\hline
Bacteria                                     & 93.825          & 97                  & 70992                                                                        & 91794                 \\ 
\hline
Dengue                                       & 99.975          & 100                 & 10676                                                                        & 174                   \\ 
\hline
\multicolumn{1}{|c|}{\textbf{Average}}       & \textbf{91.203} & \textbf{97.263}     & \multicolumn{1}{l|}{}                                                        &                       \\
\hline
\end{tabular}
}
\end{table}

Our approach outperformed the baseline approach on the benchmark datasets especially when the Median Absolute Deviation (MAD) of the sequence lengths was high. It performed quite well even on unseen classes — Bacteria and Dengue — which were not used for building UBM-GMM. This suggests that UBM-GMM captures some important characteristics of the genomic data that results in discriminative properties of the adapted mixture models of a particular class. In order to interpret these results, we tried to visualize the activations per mixture components, for sequences of a given class, during the MAP adaptation process. Figure \ref{fig:PPMC} (a) shows the active mixture components for Plants and Fungi classes along with their sub-classes. We can see from the figures that the active mixture components vary across sub-classes which provides the necessary discriminative properties that enable classification. Variations in active mixture components can also be observed across Plants and Fungi classes. Figure \ref{fig:PPMC} (b) shows the means of the most active mixture components for two sequences belonging to the Chlorophyta and Streptophyta sub-classes of the Plant class each. The means of the active mixture components are clearly distinct across the two sub-classes.

\section{Conclusion and Future Work}
\label{sec:conclusion}

We proposed a novel UBM-GMM-based subspace representation to facilitate the genomic sequence classification task. In particular, instead of length-normalizing sequences, we use a sliding window of fixed length across the sequences. The features extracted from fixed-length frames are then used to build the UBM-GMM. This helps in ensuring that the properties that vary across a sequence are indeed captured by different activations of different mixture components, and allows the representation model to generalize to genomes of unseen species as well. The next step would be to study if the ordering of these features are important for classification, especially for Fungi, where the performance is comparatively poor. A detailed phylogenetic analysis at different taxonomic levels and its correlation to the proposed methods may lead to new insights.

\bibliographystyle{IEEEbib}
\bibliography{strings,refs}

\end{document}